\documentclass[reprint,showpacs,preprintnumbers,amsmath,amssymb,aip]{revtex4-1}
\usepackage{graphicx,wasysym}
\usepackage{dcolumn}
\usepackage{bm}
\usepackage{bbm}
\usepackage{natbib}
\usepackage{tabulary}
\usepackage{color}
\usepackage{hyperref}

\usepackage{color} 
\usepackage[normalem]{ulem}

\usepackage{graphicx}
\usepackage{amsmath}
\usepackage{color}
\usepackage{float}
\usepackage{bm}

\newcommand{\tN}{t_\mathrm{N}}
\newcommand{\kB}{k_\mathrm{B}}

 \usepackage[T1]{fontenc}

\begin{document}

\title{A large deviation theory perspective on nanoscale transport phenomena}
\author{David T. Limmer} 
\email{dlimmer@berkeley.edu}
\affiliation{Department of Chemistry, University of California, Berkeley CA 94609 \looseness=-1}
\affiliation{Kavli Energy NanoScience Institute, Berkeley, CA 94609 \looseness=-1}
\affiliation{Materials Science Division, Lawrence Berkeley National Laboratory, Berkeley, CA 94609 \looseness=-1}
\affiliation{Chemical Science Division, Lawrence Berkeley National Laboratory, Berkeley, CA 94609\looseness=-1}
\author{Chloe Y. Gao} 
\affiliation{Department of Chemistry, University of California, Berkeley CA 94609 \looseness=-1}
\affiliation{Chemical Science Division, Lawrence Berkeley National Laboratory, Berkeley, CA 94609\looseness=-1}
\author{Anthony R. Poggioli} 
\affiliation{Department of Chemistry, University of California, Berkeley CA 94609 \looseness=-1}
\affiliation{Kavli Energy NanoScience Institute, Berkeley, CA 94609 \looseness=-1}

\date{\today}
\begin{abstract}
{\bf Abstract:} Understanding transport processes in complex nanoscale systems, like ionic conductivities in nanofluidic devices or heat conduction in low dimensional solids, poses the problem of examining fluctuations of currents within nonequilibrium steady states and relating those fluctuations to nonlinear or anomalous responses. We have developed a systematic framework for computing distributions of time integrated currents in molecular models and relating cumulants of those distributions to nonlinear transport coefficients. The approach elaborated upon in this perspective follows from the theory of dynamical large deviations, benefits from substantial previous formal development, and has been illustrated in several applications.  The framework provides a microscopic basis for going beyond traditional hydrodynamics in instances where local equilibrium assumptions break down, which are ubiquitous at the nanoscale. 
\end{abstract}


\maketitle

\section{Introduction}
In molecular and nanoscopic systems, fluctuations abound, material properties depend on their spatial extent, and nonlinear response is typical.  These features render the study of transport phenomena on such small scales distinct from its study on macroscopic scales, where fluctuations are suppressed and linear laws are largely valid. 
Here we review a perspective on nanoscale transport phenomena based on large deviation theory.\cite{touchette2009large} Large deviation theory provides a means of characterizing fluctuations of currents within nonequilibrium steady-states,\cite{touchette2018introduction,bertini2015macroscopic,derrida2007non} and also a practical route to evaluating the likelihood of fluctuations with computer simulations.\cite{giardina2006direct,giardina2011simulating,tchernookov2010list,ray2018importance,ray2018exact} Further, recent developments have elucidated how particular fluctuations of microscopic variables can encode nonlinear response, enabling a atomistic description of transport behavior far from equilibrium.\cite{gallavotti1996extension,shibata2002green,gaspard2013multivariate,gao2019nonlinear,barbier2018microreversibility} This enables the development of approaches that go beyond the locally linear phenomenology of traditional hydrodynamics, and allows for linear and nonlinear constitutive relations to be derived directly from molecular principles. 

The study of nanoscale transport phenomena is motivated by advances in nanofabrification techniques and experimental measurements that have driven increasingly sophisticated empirical observations into fluxes and flows on small scales. Such experimental investigations have established a number of emergent behaviors when systems are scaled down. These range from the anticipated importance of boundaries when surface-to-volume ratios are large, to unexpected violations of continuum laws valid on macroscopic scales, due to the confinement of fluctuations to low dimensions or to local departures from equilibrium.\cite{Faucher_et_al2019,Mouterde_et_al2019,Yang_et_al2018,chang2008breakdown,xu2014length,yang2010violation,wang2011non} At the same time, the continued miniaturization of devices has emphasized the importance of establishing connections between molecular properties and emergent device characteristics in order to develop novel design rules. For example, the efficiency of blue energy harvesting and waste heat storage devices depend strongly on particular chemical compositions and molecular geometries, as well as the emergent nonlinearity ubiquitous at the nanoscale.\cite{Siria_et_al2017, Laucirica_et_al2020,Lokesh_et_al2018,Zhang_et_al2015} Similarly, high throughput sensors and low power logical circuits utilize locally nonlinear responses to operate effectively and so cannot be understood with continuum theories.\cite{du2017non,wen2017generalized,gao2019nanocalorimetry,freitas2020stochastic,gao2021principles} These point to the need and timeliness of new theories bridging the divide between our traditional understanding of transport phenomena and that which occurs at the nanoscale. 

Large deviation theory has emerged as a potential formalism to fill this role, connecting nonequilibrium statistical mechanics to mesoscopic observable phenomena.  It provides anticipated scaling forms for probability distributions of time extensive random variables and their cumulant generating functions. These results underpin fluctuation theorems\cite{lebowitz1999gallavotti,gallavotti1995dynamical,kurchan1998fluctuation,crooks1999entropy,maes1999fluctuation} and thermodynamic uncertainty relations\cite{barato2015thermodynamic, gingrich2016dissipation} that provide bounds and symmetry relations for current fluctuations arbitrarily far from equilibrium. In some cases, large deviation functions also serve dual roles as generating functions, encoding statistics of currents, in addition to thermodynamic potentials, from which response relations can be derived.\cite{onsager1931reciprocal,onsager1931reciprocal2,speck2016thermodynamic,gao2019nonlinear} 
Just as  hydrodynamic theories relate transport problems near equilibrium to Gaussian fluctuations about equilibrium, large deviation theory fills a gap relating far from equilibrium,  nonlinear response ubiquitous at the nanoscale to rare fluctuations in equilibrium.
Numerical techniques to evaluate large deviation functions have been widely applied in lattice and low dimensional models of transport.\cite{jack2015hyperuniformity,hurtado2009test,espigares2013dynamical,gorissen2009density,hurtado2011spontaneous,turci2011large,speck2011space} In molecular models, similar analysis has been slower, though large deviations have recently been studied in glassy systems,\cite{hedges2009dynamic,speck2012constrained,limmer2014theory,speck2012first,pitard2011dynamic} and active matter.\cite{keta2021collective,nemoto2019optimizing,fodor2020dissipation,grandpre2021entropy} Recent advances reviewed here show that the application of large deviation theory to molecular transport models is now tractable.

In this perspective, we review some large deviation theory in the context of nanoscale transport phenomena with a focus on classical molecular systems. We illustrate how fluctuations and their associated nonlinearities can be treated consistently from a molecular, rather than phenomenological, perspective.
We first consider basic formal results, reviewing past work clarifying the structure of nonequilibrium fluctuations and their connection to response functions in principle. We then consider advances in numerical approaches that allow those formal results to be brought to bare on complex systems in practice. We illustrate some specific applications of this view on transport, where large deviation functions have been evaluated in models that provide a realistic description of physical systems.  Finally, we conclude briefly on what is needed to extend this theory of nonequilibrium steady-states to more general classes of systems, and discuss open areas worthy of pursuit. 

\section{Large deviations in principle}
A central problem in nonequilibrium statistical mechanics is the evaluation of the probability distribution of fluctuating variables. Large deviation theory provides a path to do this. 
In the context of transport and dynamical systems, the relevant stochastic variable whose extent can be take arbitrarily large is a time integrated current, $J$, or generalized displacement,
\begin{equation}
J = \int_0^{\tN} dt \,  \dot{J}(t)
\end{equation}
where $\dot{J}(t)$ is an instantaneous flux at time $t$, and $\tN$ is the observation time, taken to be larger than any characteristic correlation time of $\dot{J}(t)$. 

The fluctuations of $J$ can be characterized by a probability distribution, or alternatively by its characteristic function. For a dynamical property or a system away from equilibrium where Boltzmann statistics do not hold, the calculation of either is a daunting task. The fundamental principle of large deviation theory is that currents that are correlated over a finite amount of time admit an asymptotic, time intensive form of the logarithm of their distribution function\cite{derrida2007non,touchette2018introduction}
\begin{equation}
\label{Eq:Phi}
\phi_E(j) = \lim_{\tN \rightarrow \infty} \frac{1}{\tN} \ln \langle \delta (j \tN-{J}[X]) \rangle_E 
\end{equation}
where $\phi_E(j)$ is known as the rate function and is a natural variable of the time intensive current, $j=J/\tN$. We will adopt a notation that distinguishes averages in the presence of an external field $E$ that drives the current, where $\langle .. \rangle_E$ denotes an average in the steady state generated by field $E$,  and $\delta(j \tN-J[X])$ is Dirac's delta function evaluated using a fluctuating current ${J}[X]$ that depends on a trajectory $X$. This asymptotic form implies that deviations away from the mean are exponentially unlikely, with a rate set by $\phi_E(j)$. From large deviation theory, the characteristic function associated with fluctuations of $J$ can be computed from the Laplace transform of Eq.~\ref{Eq:Phi},
\begin{equation}
\label{Eq:Psi}
\psi_E(\lambda)= \lim_{\tN \rightarrow \infty}  \frac{1}{\tN}\ln \left \langle e^{- \lambda J} \right \rangle_E
\end{equation}
where $\psi_E(\lambda)$ is known as a scaled cumulant generating function, and depends on the Laplace parameter $\lambda$ but not $\tN$. Derivatives of $\psi_E(\lambda)$ with respect to $\lambda$ evaluated at $\lambda=0$ yield the time intensive cumulants of $J$. 

\begin{figure}
\begin{center}
\includegraphics[width=8.5cm]{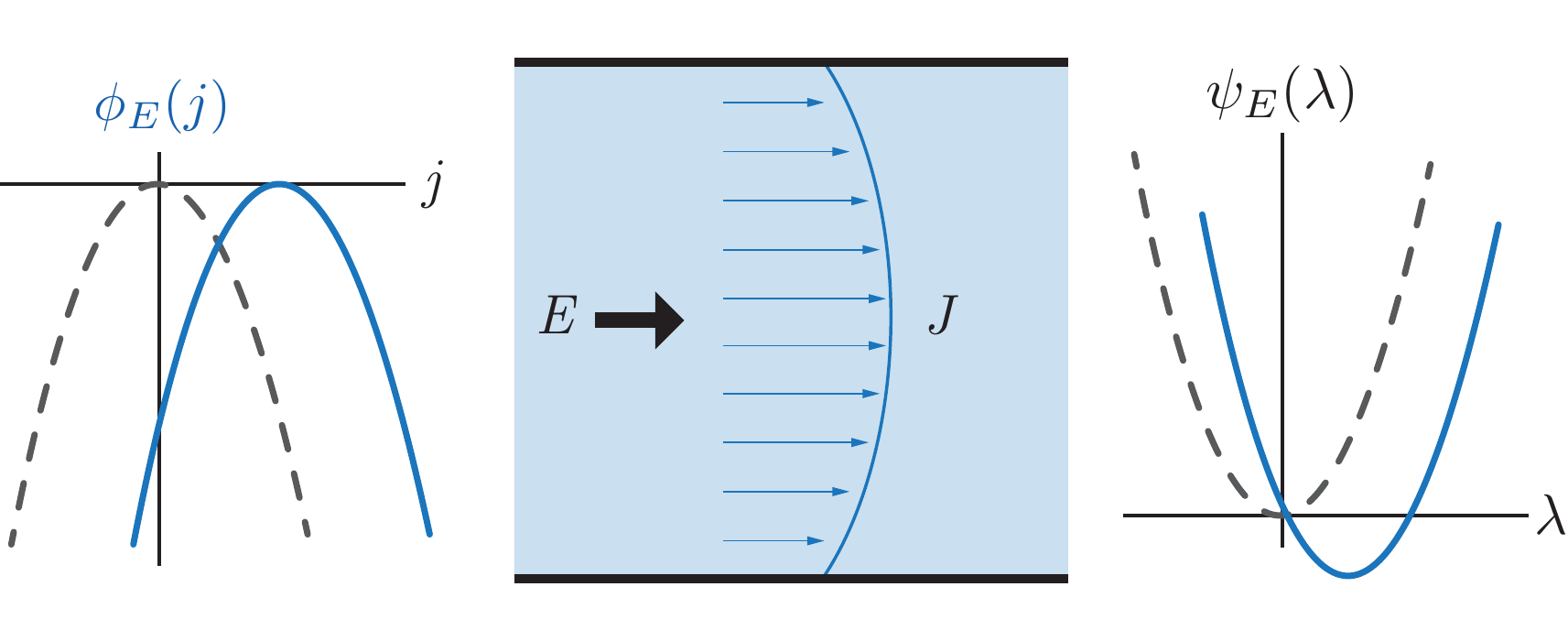}
\caption{Illustration of the expected form of the rate function, $\phi_E(j)$, and cumulant generating function, $\psi_E(\lambda)$, under small applied field $E$ (solid lines) and at equilibrium (dashed lines).}
\label{Fi:0}
\end{center} 
\end{figure}

The pair $\phi_E(j)$ and $\psi_E(\lambda)$ are related to partition functions for path ensembles that are either conditioned to exhibit a value of $J$ or that are statistically biased towards different $J$'s through $\lambda$.\cite{chandler2010dynamics,jack2020ergodicity} They completely characterize the fluctuations of the current within a nonequilibrium steady-state. 
When both $\phi_E(j)$ and $\psi_E(\lambda)$ exist and are smooth and differentiable, they contain the same information. Specifically, they are related to each other through a Legendre-Fenchel transform,\cite{touchette2009large}
\begin{eqnarray}
\label{Eq:LFT}
\phi_E(j)  &=&  \inf_\lambda [ \psi_E(\lambda)+\lambda j ] 
\end{eqnarray}
as follows from a saddle point approximation to the integral definition, valid in the long time limit. Further, when this relation holds, trajectories taken from the conditioned ensemble are equivalent to trajectories under the statistical bias.\cite{chetrite2013nonequilibrium} Figure~\ref{Fi:0} illustrates these two quantities near equilibrium.

Many formal results have been derived for large deviation functions of time integrated currents. Some of the most foundational and useful in the context of nanoscale transport are reviewed below. In the following, a distinction is drawn between systems near equilibrium and those far from it.  

\subsection{Fluctuations near equilibrium}
Assuming that the departure from equilibrium is small and that Boltzmann statistics hold, Onsager originally conjectured that the log-probability of a current fluctuation was given by the entropy production to create it.\cite{onsager1931reciprocal,onsager1931reciprocal2} In what he called a dissipation function, now identifiable as a rate function, the log-probability of a current fluctuation was given by
\begin{eqnarray}
\phi_0(j) &\approx& -\beta E j /4 \\
&=&  -\beta j^2/4  \chi \nonumber
\end{eqnarray}
where in the second line the dependence on the field has been eliminated through a phenomenological linear law relating the current to the field, $j = \chi E $, through a constant of proportionality $\chi$ that is observed to be a generalized conductivity. The Gaussian form for the current fluctuations is consistent with microscopic reversibility in equilibrium that requires $\phi_0(j)$ is an even function of $j$. 

The corresponding scaled cumulant generating function is, 
%choosing the counting variable to be $\beta E$, is
\begin{eqnarray}
\psi_0( \lambda) &=& \lambda^2 \chi/ \beta
\end{eqnarray}
where under the Gaussian approximation to the current fluctuations, $\psi_0(\lambda)$ is quadratic in $\lambda$. As a scaled cumulant generating function, derivatives of $\psi_0( \lambda)$ with respect to its argument provide the intensive cumulants of $J$, for example
%%and for small departures from equilibrium
%\begin{eqnarray}
%\frac{d \psi_0(\lambda)}{d \lambda}  &=& -\langle J \rangle_0/\tN  \\
%&=& 2 \lambda \chi /\beta \nonumber
%\end{eqnarray}
%and
%we recover the linear law imposed in the rate function. This implies that near equilibrium $\psi( \beta E)$ acts as a thermodynamic potential, suggesting $J$ and $E$ are conjugate quantities. An immediate consequence of this form of $\psi(\lambda)$ is the fluctuation dissipation relation 
\begin{eqnarray}
\label{Eq:LR}
\left . \frac{d^2 \psi_0(\lambda)}{d \lambda^2 } \right |_{\lambda=0}&=& \frac{1}{\tN}\langle \delta J^2 \rangle_0\\
&=& 2\chi/ \beta  \nonumber %\beta  \langle \delta J^2 \rangle_0 \nonumber 
\end{eqnarray}
which relates the conductivity with the variance of the current, $\chi = \beta \langle \delta J^2 \rangle_0 /2 \tN$. In the long time limit this is equivalent to
\begin{eqnarray}
\label{Eq8}
\chi &=& \beta  \int_0^{\tN} dt \, \langle j(0) j(t) \rangle_0
\end{eqnarray}
assuming current correlations decay faster than $1/t$. The first form of the fluctuation dissipation relation is known as an Einstein-Hefland moment.\cite{helfand1960transport,viscardy2007transport} Equation \ref{Eq8} follows from time reversal symmetry, and results in a traditional Green-Kubo expression for the response of a current in terms of an integrated time correlation function.\cite{green1951brownian,green1954markoff,zwanzig1965time}

The Gaussian form of $\phi_0(j)$ is valid only for small $j$, as is the subsequent linear response relationship that follows. Their utility derives from their thermodynamic origin, in which the entropy production uniquely determines the response. This endows linear response relationships with great generality. They are equally valid independent of the specific dynamics of the system, provided the large deviation form holds. In practice, this requires that correlation times for the current are finite.  %Traditional complications encountered in arriving at this Green-Kubo relationship are avoided under this large deviation scaling assumption, which requires all cumulants of $J$ to scale linearly with $\tN$.\cite{zwanzig1964elementary}

\subsection{Fluctuations far from equilibrium}
Unlike fluctuations about an equilibrium steady-state, fluctuations away from equilibrium are not generically determined solely by thermodynamic considerations.\cite{maes2018non} Thermodynamics can bound the scale of fluctuations and impose specific symmetries, but their detailed form will depend in general on the specific equations of motion developing those fluctuations. In the following we will restrict our discussion to processes describable by a Langevin equation.\cite{zwanzig2001nonequilibrium} For concreteness, in the next few sections, we will consider an underdamped equation of the form
\begin{equation}
\label{Eq:EOM}
\dot{\mathbf{r}} = {\bf{v}}_i \, ,\quad m_i \dot{\bf{v}}_i = \mathbf{F}_i \left [\mathbf{r}^N \right ] + \mathbf{E}_i - \gamma_i \mathbf{v}_i +\boldsymbol{\eta}_i 
\end{equation}
for particle $i$ where the final two terms obey a local detailed balance
 with a temperature $T_i$, by dissipating energy through the friction, $\gamma_i$, and adding energy by a random force $\boldsymbol{\eta}_i(t)$ with Gaussian statistics described by $\langle \boldsymbol{\eta}_i(t) \rangle = 0$, $\langle \boldsymbol{\eta}_i(t)\boldsymbol{\eta}^{T}_j(t') \rangle = 2 \gamma_i \kB T_i \mathbf{1} \delta_{ij}\delta(t-t')$, where $\kB$ is Boltzmann's constant and $\mathbf{1}$ is the unit matrix. The force, $\mathbf{F}_i \left [\mathbf{r}^N \right ]$, is assumed to be  gradient but depends on the full configuration of the system, $\mathbf{r}^N$, and $\mathbf{E}_i$ is an external field driving a nonequilibrium steady-state. In the limit that $\mathbf{E}_i =0$ and $T_i = T$ for all $i$, the system evolving with Eq.~\ref{Eq:EOM} will develop a Boltzmann distributed steady state and exhibit microscopic reversibility.\cite{crooks2011thermodynamic} 
 
 A consequence of the underlying microscopic reversibility of Eq.~\ref{Eq:EOM}, and its local detailed balance, is that when driven away from equilibrium its trajectories satisfy the Crooks fluctuation theorem.\cite{crooks1999entropy} In terms of a scalar current $j$ driven by a scalar field $E$, this symmetry is manifest in the current rate function as
\begin{equation}
\label{Eq:FT}
\phi_E(j) - \phi_E(-j) = \beta E j 
\end{equation}
due to Kurchan for diffusive dynamics.\cite{kurchan1998fluctuation} This symmetry means that currents that evolve in opposition to their driving field are exponentially unlikely with a scale set by the entropy production associated with the current.  This fluctuation theorem is a specific realization of a more general fluctuation theorem for the total entropy production,\cite{seifert2005entropy} can be generalized to multiple currents,\cite{gaspard2013multivariate} and is a microscopic statement of the second law of thermodynamics.\cite{seifert2012stochastic} Importantly this relationship is valid for arbitrary $E$. When $E=0$, it reduces to a condition that the equilibrium probability of a current is an even function, from which the linear response relationships discussed above follow.\cite{searles2000fluctuation} Its equivalent statement in terms of the scaled cumulant generating function is
\begin{equation}
\label{Eq:LSGC}
\psi_E(\lambda) = \psi_E(\beta E-\lambda)
\end{equation}
which in the context of stochastic dynamics is known as Lebowitz-Spohn symmetry,\cite{lebowitz1999gallavotti} or in the deterministic limit as Gallavati-Cohen symmetry.\cite{gallavotti1995dynamical} As these symmetry relations are thermodynamic in origin, equivalent expressions exist independent of the specific evolution equation. 

Around an equilibrium steady-state, Eq.~\ref{Eq:FT} or Eq.~\ref{Eq:LSGC}, is sufficient to deduce a linear response relationship in the form of the fluctuation-dissipation theorem.\cite{chandler1987introduction} This reflects the fact that, in this specific case, the fluctuations of $J$ in equilibrium encode the response of $J$ to the field $E$. However, for nonlinear response, or linear response around a nonequilibrium steady-state, additional functional information is needed. One way to proceed elaborated upon by Gaspard \cite{gaspard2013multivariate} evaluates the average current $\langle J \rangle_E$ in terms of mixed derivatives of $\psi_E(\lambda)$. Using the fluctuation theorem, the current up to second order in the field is
\begin{eqnarray}
\label{Eq:Gaspard}
\frac{\langle J \rangle_E}{\tN} &=& \left . \frac{\partial^2 \psi_0(\lambda)}{\partial \lambda^2} \right |_{\lambda=0} \frac{\beta E}{2} +\left . \frac{\partial^3 \psi_E(\lambda)}{\partial \lambda^2 \partial \beta E} \right |_{\lambda,E=0} \frac{(\beta E)^2}{2} \nonumber \\
&=& \frac{\langle \delta J^2 \rangle_0}{2 \tN} \beta E +\frac{1}{2 \tN} \left . \frac{d \langle \delta J^2 \rangle_E}{d \beta E}  \right |_{E=0} (\beta E)^2
\end{eqnarray}
where the second line follows from the definition of the scaled cumulant generating function.\cite{gaspard2013multivariate,barbier2018microreversibility} This illustrates that in addition to knowledge of the fluctuations of $J$ about equilibrium, it is necessary to know how those fluctuations change with an applied field to predict higher order response. This is due to the general breakdown of the fluctuation-dissipation relationship away from equilibrium, and the fact that kinetic factors, like the effective diffusivity $\langle \delta J^2 \rangle_E$, become important within nonequilibrium steady-states. 

An alternative way to interpret the fact that equilibrium fluctuations of $J$ are insufficient to predict the full response of a current to a field $E$ is to note that only near equilibrium are $J$ and $E$ conjugate dynamical quantities. From Maes and coworkers, away from equilibrium, in general the quantity conjugate to $E$ in the path probability will have components that are asymmetric under time reversal, as well as components that are symmetric under time reversal.\cite{baiesi2013update, baiesi2010nonequilibrium} While the fluctuation theorem uniquely determines the former to be the entropy production, the specific form of the latter depends on the equation of motion. If we denote $Q$ as the fluctuating time reversal symmetric contribution conjugate to $E$ in the path probability, which is zero on average in equilibrium, then 
\begin{equation}
Q = \int_0^{\tN} dt \, \dot{Q}(t)
\end{equation}
which is extensive in time with increment $\dot{Q}$ and $q=Q/\tN$ its intensive counterpart. Considering the joint rate function, $\hat{\phi}_E(j,q)$ for the time intensive $j$ and $q$, we have from the fluctuation theorem
\begin{equation}
\hat{\phi}_E(j,q) - \hat{\phi}_E(-j,q) = \beta E j
\end{equation}
while its complement
\begin{equation}
\hat{\phi}_E(j,q) + \hat{\phi}_E(-j,q) = \beta D_E(q)
\end{equation}
defines a function of $q$ that encodes the time symmetric contribution, $\beta D_E(q)$, to the joint rate function $\hat{\phi}_E(j,q)$. This function can be readily calculated from an explicit equation of motion like that in Eq.~\ref{Eq:EOM}, and specific forms are shown in Sec.~\ref{Sec:ED} and Sec.~\ref{Sec:NanoF}. Using these two relations, both valid for arbitrary $E$, we can relate the joint rate function $\hat{\phi}_E(j,q)$ under the applied field to its value in the absence of the field, $\hat{\phi}_0(j,q)$, as
\begin{equation}
\hat{\phi}_E(j,q) -\hat{\phi}_0(j,q) = \beta  E j/2  + \beta \Delta D_E(q)/2
\end{equation}
where $\Delta D_E(q) = D_E(q)-D_0(q)$ is referred to as the excess dynamical activity, the time symmetric analogue to the entropy production. 

\begin{figure}
\begin{center}
\includegraphics[width=8.5cm]{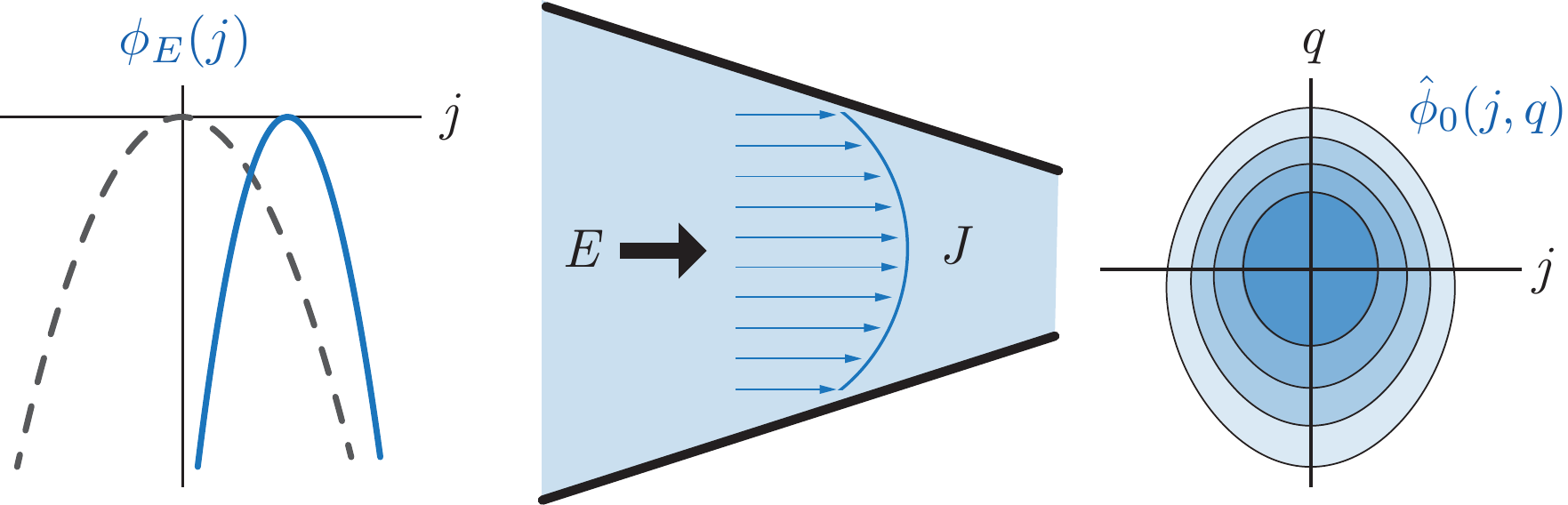}
\caption{Nonlinear response like current rectification through a diode can be understood by how (left) current fluctuations described by $\phi_E(j)$ change  at equilibrium (dashed lines) or for large fields (solid lines), or alternatively how (right) current and activity fluctuations described by $\hat{\phi}_0(j,q)$ are correlated in equilibrium.}
\label{Fi:00}
\end{center} 
\end{figure}
While for general dynamical processes and arbitrary currents $\Delta D_E(q)$ may be complicated, for currents that are linear in the microscopic velocities driven by fields that enter linearly into the equation of motion given in Eq.~\ref{Eq:EOM}, $\Delta D_E(q)$ simplifies significantly. Specifically, we have found that it can be deduced to be linear in $q$ with an additive constant proportional to $E^2$.\cite{gao2019nonlinear} This means that the joint rate function for $j$ and $q$, driven arbitrarily far from equilibrium by $E$, is related to its equilibrium counterpart by
\begin{equation}
\label{Eq:Gao}
\hat{\phi}_E(j,q) -\hat{\phi}_0(j,q) = \beta  E (j+q)/2 - \beta  \chi_\mathrm{id} E^2/4
\end{equation}
where $\chi_\mathrm{id}$ is the conductivity in the non-interacting particle limit.  Analogously, the scaled cumulant generating function is
\begin{eqnarray}
\label{Eq:Gao2}
\hat{\psi}_E(\lambda_j,\lambda_q) &=& \frac{1}{\tN} \ln \left \langle e^{-\lambda_j J-\lambda_q Q} \right \rangle_E \\
&=& \hat{\psi}_0(\lambda_j - \beta E/2,\lambda_q - \beta E/2) -  \beta  \chi_\mathrm{id} E^2/4 \nonumber
\end{eqnarray}%
This implies that, for systems in which Eqs.~\ref{Eq:Gao} and \ref{Eq:Gao2} hold, $\hat{\phi}_0(j,q)$ acts as a thermodynamic potential that completely determines the response of a current to an applied field.\cite{gao2019nonlinear} Further, it implies a nonequilibrium ensemble reweighting principle between steady-states evolved under different applied fields with $E$ and the sum $J+Q$ acting as conjugate dynamical variables.\cite{lesnicki2020field} While this is restricted to diffusions of the form of Eq.~\ref{Eq:EOM}, a similar result has recently been proved for driven exclusion processes.\cite{vanicat2020mapping} Using these relations,  the average current $\langle J \rangle_E$ is
\begin{eqnarray}
\label{Eq:GaoJ}
\langle J \rangle_E &=& \langle J e^{\beta  E (J+Q)/2 -\beta  \chi_\mathrm{id} \tN E^2/4} \rangle_0
\end{eqnarray}
valid for arbitrary $E$. To second order in the field, the current is given by
\begin{eqnarray}
\label{Eq:Gao3}
\frac{\langle J \rangle_E}{\tN} &=&\left . \frac{\partial^2 \hat{\psi}_0(\lambda_j,\lambda_q) }{\partial \lambda_j^2}\right |_{\lambda=0} \frac{\beta E}{2} - \left .\frac{\partial^3 \hat{\psi}_0(\lambda_j,\lambda_q) }{\partial \lambda_j^2 \partial \lambda_q} \right |_{\lambda=0}\frac{(\beta E)^2}{4} \nonumber \\
&=& \frac{\langle \delta J^2 \rangle_0}{\tN} \frac{\beta E}{2}   + \frac{\langle \delta J^2 \delta Q \rangle_0}{\tN} \frac{(\beta E)^2}{4}
\end{eqnarray}
where we have invoked the time reversal symmetry of the equilibrium average to eliminate terms that average to 0 and employed the fact that $\langle J \rangle_0 = \langle Q \rangle_0=0$.\cite{gao2019nonlinear} Comparing to Eq.~\ref{Eq:Gaspard}, we observe that the correlations between $J$ and $Q$, $\langle \delta J^2 \delta Q \rangle_0$, encode the change of the fluctuations in $J$ with the applied field through an equilibrium expectation value. These two views are illustrated in Fig.~\ref{Fi:00}. Thus, to construct response theories for nanoscale transport where nonlinearities and departures from equilibrium are commonplace, one can either determine the functional dependence of the distribution of currents on the driving field, or consider the joint distribution of the current and activity in equilibrium. 

%\subsection{Connections to previous perspectives}
%--kawaski/evans/searls
%--MFT

\section{Large deviations in practice}
%Large deviation functions characterize fluctuations and encode nonlinear current-field relationships. 
While large deviation functions have been evaluated in analytically tractable systems, and in lattice models, the application of large deviation theory to nanoscale transport problems with detailed molecular or mesoscopic models has been limited. In large part this is due to the lack of suitable numerical tools.  With recent advances by us and others, this formalism can now be brought to bare on complex systems. Below we review some existing techniques for evaluating large deviation functions numerically. We distinguish two existing approaches that either attempt to solve for a large deviation function directly by solving a generalized eigenvalue equation from those that are based on estimating them stochastically by sampling rare trajectories.  

\subsection{Evaluating large deviation functions directly}
\label{Sec:ED}
The traditional approach to compute the large deviation functions is to employ the Feynman-Kac theorem,\cite{majumdar2007brownian,touchette2018introduction,chetrite2013nonequilibrium} which relates the scaled cumulant generating function to the largest eigenvalue of a tilted or deformed operator.\cite{lebowitz1999gallavotti} For the Markovian process in Eq.~\ref{Eq:EOM}, and an observable of the form $j=\sum_i \mathbf{a}_i \cdot \mathbf{\dot{r}}_i+\mathbf{b}_i \cdot \dot{\mathbf{v}}_i$, the generalized eigenvalue equation is
\begin{equation}
\label{Eq:Ltilt}
\mathcal{L}_\lambda \mathcal{R}_\lambda = \psi_E(\lambda) \mathcal{R}_\lambda
\end{equation}
and 
\begin{eqnarray}
\mathcal{L}_\lambda =\sum_i && \mathbf{v}_i \nabla_{\mathbf{r}_i} + \frac{1}{m_i} [\mathbf{F}_i(\mathbf{r}^N)+\mathbf{E}_i]\left ( \nabla_{\mathbf{v}_i}  + \lambda \mathbf{b}_i \right ) \nonumber \\
&+& \frac{\kB T_i \gamma_i}{m_i^2} \left ( \nabla_{\mathbf{v}_i} +  \lambda \mathbf{b}_i \right )^2 - \lambda \mathbf{a}_i  \mathbf{v}_i
\end{eqnarray}
where the adjoint of $\mathcal{L}_0$ is the Fokker-Planck operator and $\mathcal{R}_\lambda$ is the dominate right eigenvector. 
%A distinction is made between the left and right vector spaces associated with $\mathcal{L}_\lambda$, because it is not in general Hermitian. 
%In fact, generally both Eq.~\ref{Eq:Ltilt} and its adjoint need to be solved simultaneously to compute $\psi_E(\lambda)$. Despite being non-Hermitian,  whenever the spectrum of $\mathcal{L}_\lambda$ is gapped, the Perron-Frobenius Theorem guarantees that $\psi_E(\lambda)$ is real, and normalization of the probability distribution sets $\psi_E(0)=0$.\cite{}
%
Typically the force $F_i(\mathbf{r}^N)$ complicates the analytic solution of the eigenvalue equation. In force-free cases, like free diffusion\cite{derrida2005einstein} and open Levy walks,\cite{dhar2013exact} and linear systems, like collections of harmonic oscillators\cite{kundu2011large} and Ornstein-Uhlenbeck processes,\cite{jack2020ergodicity,chetrite2015nonequilibrium,fogedby2011bound} the cumulant generating function can be calculated. In Sec.~\ref{Sec:ABP} we review a study on the diffusive transport of a tagged active Brownian particle, which can be solved exactly by integrating out degrees of freedom with a many body expansion.\cite{grandpre2018current} 

\begin{figure}
\begin{center}
\includegraphics[width=8.5cm]{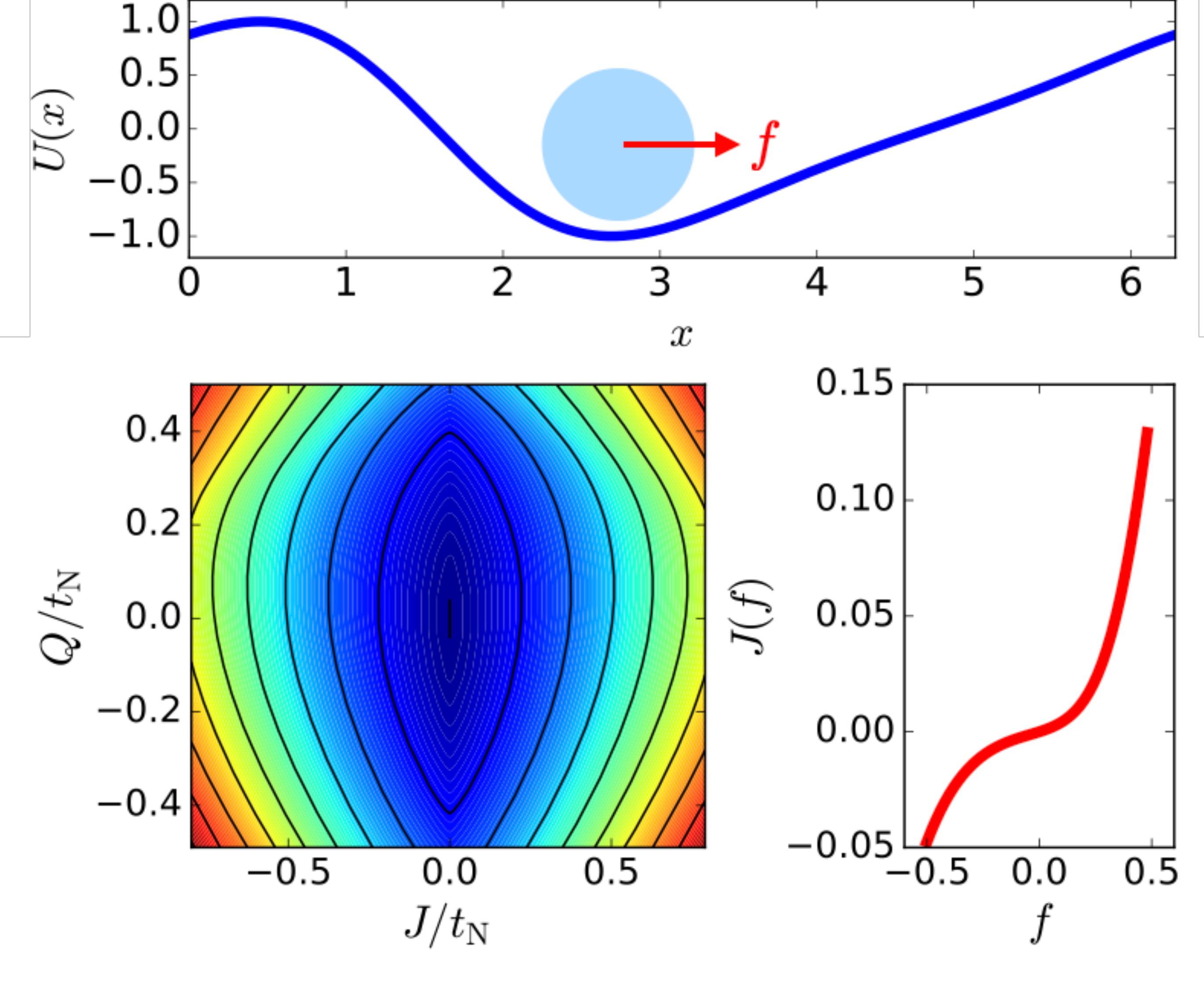}
\caption{(top) Illustration of a Brownian particle in a ratchet potential subject to an applied force. (Bottom left)  The joint rate function $\hat{\phi}_0(j,q)$ encoding the response of the particle's current. (Bottom right) The associated current as a function of the applied drift force computable from $\hat{\phi}_0(j,q)$ using Eq.~\ref{Eq:GaoJ}.}
\label{Fi:2}
\end{center} 
\end{figure}

Absent analytical solutions, basis set techniques can be used to numerically solve Eq.~\ref{Eq:Ltilt}. For example Fig. ~\ref{Fi:2} illustrates the joint rate function $\hat{\phi}_0(j,q)$ for diffusive transport in a periodic ratchet potential. An overdamped equation of motion reduces the dimensionality of the system to a simple periodic coordinate, $r$,
\begin{equation}
\gamma \dot{r} =  F(r) + f + \eta \quad 0 \leq r \leq 2\pi
\end{equation}
and $\hat{\phi}_0(j,q)$ is the joint rate function for $j=\dot{r}$ and corresponding dynamical activity  $q=-\gamma^{-1} F(r)$, the negative external force. As a periodic problem, Eq.~\ref{Eq:Ltilt} can be diagonalized using a Fourier basis and $\hat{\phi}_0(j,q)$ evaluated using an analogue of Eq.~\ref{Eq:LFT}. In this specific system, $F(r) = -\partial_r U(r)$, where $U(r)= -\sin(r + \sin(r)/2)$, lacks inversion symmetry, and the nonlinear response of $j$ to an applied force $f$ exhibits current rectification, see Fig. ~\ref{Fi:2}. The joint $j$ and $q$ fluctuations encode this asymmetric response. In the joint rate function, fluctuations that increase $q$ are correlated with increasing the scale of fluctuations in $j$, or $\langle \delta J^2 \delta Q \rangle_0>0$, and vice versa, leading to a larger current and differential mobility for $f>0$ than for $f<0$, from Eq. \ref{Eq:Gao3}.  For interacting problems, significant developments into compact basis sets employing matrix and tensor product states have advanced the state of the art for lattice transport problems.\cite{banuls2019using,causer2021optimal,helms2020dynamical,helms2019dynamical,johnson2010dynamical} However, these have not yet been translated into the continuum for molecular problems. 

A very attractive alternative to the generalized eigenvalue equation for the direct evaluation of the large deviation functions is through variational principles that  $\psi_E(\lambda)$ satisfies.\cite{nemoto2011thermodynamic,chetrite2015variational,das2019variational,ray2020constructing,jacobson2019direct} While the non-Hermitian nature of $\mathcal{L}_\lambda$ precludes traditional eigenvalue-based variational statements from being formulated, a result from optimal control theory can be used. Specifically, employing an auxiliary system with additional drift $\mathbf{u}(\mathbf{r}^N,\mathbf{v}^N)$
\begin{equation}
\label{Eq:Var}
\psi_E(\lambda) = \sup_{\mathbf{u}(\mathbf{r}^N,\mathbf{v}^N)} \left [-\lambda \langle j  \rangle_{u} +   \langle \Delta \mathcal{U}  \rangle_{u} \right ]
\end{equation}
where 
\begin{equation}
\Delta \mathcal{U}  = \sum_i \frac{\mathbf{u}}{4 \kB T_i \gamma_i}  \left (\mathbf{u}-2 m \dot{\mathbf{v}}_i+2 \mathbf{F}_i+\mathbf{E}_i -\gamma_i \mathbf{v} \right )
\end{equation}
is the change in the action following from a Girsanov transform and application of Jensen's inequality. Here $\langle \dots \rangle_{u}$ denotes an average with the additional drift. The maximization is over all control drifts and is solved by the dominant eigenvector with component $u_i(\mathbf{r}^N,\mathbf{v}^N) = 2 \kB T_i/\gamma_i \nabla_i \ln \mathcal{R}_\lambda$.\cite{jack2010large,jack2015effective,chetrite2013nonequilibrium} While the eigenvector is many-bodied, low rank approximate forms have been optimized using analogues of variational Monte Carlo\cite{das2019variational} and machine learning,\cite{rose2021reinforcement,whitelam2020evolutionary} which have been found to yield accurate results. The variational Monte Carlo approach is made efficient by explicit forms for the derivatives of Eq.~\ref{Eq:Var} with respect to the added drift.\cite{das2019variational} Further, perturbative corrections in the form of a cumulant expansion can be formulated to increase the accuracy of the estimate at the cost of breaking the variational structure.\cite{das2019variational,klymko2018rare} This method has been applied to low dimensional models and colloidal assemblies in shear flows.\cite{das2021variational} With the concurrent development of expressive forms for $\mathbf{u}(\mathbf{r}^N,\mathbf{v}^N)$, this technique is poised to be widely applied to molecular systems.

\subsection{Estimating large deviation functions stochastically}
\label{Sec:DMC}
An alternative to the direct evaluation of a large deviation function is to estimate it by sampling molecular dynamics simulations. This direction has seen significant recent development as a means of avoiding nonequilibrium simulations that induce long range correlations or  to evaluate field-dependent differential transport coefficients directly. 
In order to accurately estimate the large deviation function, one must sample a path ensemble that incorporates the exponentially rare fluctuations in the dynamical observable of interest.
  If $P_E[X]$ denotes a reference path ensemble driven by a field $E$, then the path ensemble to be sampled to compute $\psi_E(\lambda)$ or $\phi_E(j)$ for a current $J$ is
\begin{equation}
\label{Eq:PfTilt}
P_{\lambda}[X] = P_E[X] e^{-\lambda J[X] - \psi(\lambda) \tN}
\end{equation}
where the new path ensemble $P_{\lambda}[X]$ is biased by the factor $\exp[-\lambda J ]$ and normalized by $\psi(\lambda)$. While the dynamics that generates $P_E[X]$ are defined by the model, the dynamics that generate $P_{\lambda}[X] $ are determined by the solution of the generalized eigenvalue problem in Eq.~\ref{Eq:Ltilt}. Therefore, the weight factor $\exp[-\lambda J]$ is typically incorporated through an importance sampling process on top of direct dynamical propagation. 

Two main classes of trajectory importance sampling exist, transition path sampling\cite{bolhuis2002transition} and diffusion Monte Carlo, or the cloning algorithm.\cite{tchernookov2010list,giardina2011simulating,giardina2006direct} Transition path sampling performs a sequential update to a single trajectory with fixed time in the manner of a Markov chain Monte Carlo algorithm, though through the trajectory space.\cite{dellago1998transition} The weighting factor is accommodated by a trajectory acceptance criteria. The application of transition path sampling to large deviation functions was first applied in the context of equilibrium glass formation problems,\cite{merolle2005space,hedges2009dynamic} and later extended to transport problems evolving non-detailed balanced dynamics.\cite{jack2015hyperuniformity,ray2018importance} Alternatively, the cloning algorithm propagates an ensemble of short trajectories in parallel using the reference dynamics. Each trajectory accumulates a local weight which is used as a basis for a population dynamics that reduce the variance of the weights by a branching and annihilation process. The cloning algorithm has been used widely in model transport problems.\cite{hurtado2009test,espigares2013dynamical,gorissen2009density,hurtado2011spontaneous,turci2011large}

Both transition path sampling and the cloning algorithm can evaluate large deviation functions sufficiently accurately to be used to compute transport coefficients. 
In particular, in recent work, we showed that the use of the cloning algorithm can yield statistically superior estimates of linear transport coefficients using the direct calculation of the curvature of the large deviation function through Eq.~\ref{Eq:LR}, as compared to their direct evaluation through Green-Kubo theory.\cite{gao2017transport}
This was demonstrated in the calculation of the thermal conductivity in a WCA solid and the liquid-solid friction in a Lennard Jones solution. In Sec.~\ref{Sec:CNT} we show how this procedure was used to study the anomalous heat transport in low dimensional carbon lattices.\cite{ray2019heat} The cloning algorithm has also been used to evaluate the joint large deviation function of the current and activity, and efficiently estimate nonlinear response functions by Eqs. \ref{Eq:Gao2} and \ref{Eq:Gao3}. For example, we have analyzed the rectification of heat currents in chains of nonlinear oscillators with inhomogeneous mass distributions.\cite{gao2019nonlinear} 

While the calculation of linear transport properties are tractable even for models with complex forcefields, the evaluation of nonlinear response is significantly more computationally challenging. This is because the cloning algorithm, as well as transition path sampling, both augment the propagation of the bare system dynamics with importance sampling, without any guidance from the rare events that contribute to the large deviation function. As a consequence, for exponentially rarer fluctuations, both Monte Carlo algorithms require exponentially more samples of the targeted stationary distribution as the overlap between it and the proposed distribution becomes exponentially small.\cite{hidalgo2017finite,ray2018importance,ray2018exact} Recent advances that incorporate auxiliary dynamics to guide the path sampling using approximate solutions to Eq.~\ref{Eq:Var} have been successful at significantly dropping the computational cost of both algorithms. In Ref.~\onlinecite{das2019variational}  we report a gain of over 3 orders of magnitude in the statistical efficiency of the cloning algorithm with a guiding force. Various approaches have incorporated analytical approximations,\cite{ray2018exact,grandpre2021entropy} variationally optimized ansatzes,\cite{das2019variational,ray2020constructing} or feedback control procedures.\cite{nemoto2016population,nemoto2017finite,brewer2018efficient}

An alternative route we have pursued for the evaluation of nonlinear current-field relationships is to leverage the reweighting principle valid between equilibrium and nonequilibrium ensembles when Eq.~\ref{Eq:Gao} holds.  In such a case, the probability of observing a trajectory in a driven ensemble is equal to the probability of that trajectory in equilibrium times a weighting factor
\begin{equation}
\label{Eq:PfTilt}
 P_0[X] =P_E[X] e^{-\beta (J +Q)E/2 +\beta  \chi_\mathrm{id} E^2/4}
\end{equation}
where the sum $J +Q$ depends on the trajectory, but $\beta  \chi_\mathrm{id} E^2/4$ is a trajectory independent constant. Using a series of different nonequilibrium ensembles at different values of $E$, rare equilibrium fluctuations of the current and activity can be probed.\cite{lesnicki2020field, lesnicki2021molecular} Employing standard equilibrium techniques like the weighted histogram analysis method\cite{kumar1992weighted} or multistate Bennet acceptance ratio,\cite{shirts2008statistically} many simulations can be combined to enhance expectation values at each field. Analogous to standard histogram reweighting methods used between equilibrium ensembles,\cite{frenkel2001understanding} this approach offers an attractive ability to compute the average current as a continuous function of the field using Eq.~\ref{Eq:GaoJ}. Formally, this bypasses assumptions of the existence of a Taylor series expansion of the current in terms of the applied field. Practically, it avoids having to numerically evaluate the gradient as would be necessary in direct nonequilibrium simulations. In Sec. ~\ref{Sec:NanoF} we illustrate how this procedure is used to study the field dependence of the ionic conductivity in electrolyte solutions.
 
\section{applications}
Building upon the formal developments connecting large deviation theory and nanoscale transport, and the recent advances in numerical techniques to characterize dynamical fluctuations in complex systems, a number of different canonical transport problems have bene studied. In the following, we illustrate a few specific examples in which the currents result from single tagged particles or their collections. We consider the examples of the transport of mass, energy and charge. In each, large deviation theory enables the computational study of either linear or nonlinear phenomena, in a way not feasible without the tools it provides. 

\subsection{Active brownian particle diffusion} 
\label{Sec:ABP}
Much of the framework outlined above is agnostic to whether the system is evolving within an equilibrium or nonequilibrium steady-state. As an illustration of the latter, the diffusion of a tagged particle in an active fluid is considered.\cite{grandpre2018current} An active fluid is one in which a constant source of energy is continuously converted into directed motion of individual particles, and can be realized with synthetic colloids or swimming bacteria.
%\cite{Hill2007,Lauga2006,Fu2012,Parsek2005,Palacci2013,Narayan2006,Howse2007,Walther2008,
\cite{Hill2007,Lauga2006,Fu2012b,Parsek2005b,Palacci2013,Narayan2006b,Howse2007b,Walther2008b,Bricard2013b,palacci2013living} In such a fluid, novel transport processes disallowed in equilibrium can occur due to the time-reversal symmetry breaking inherited from the persistent dissipation.\cite{hargus2021odd, wagner2019response,stenhammar2016light,reichhardt2017ratchet} For example, the viscosity of the fluid can acquire odd components, and the Stokes-Einstein relation can break down.\cite{solon2015active,banerjee2017odd,dal2019linear} 

The specific active fluid previously considered was a collection of active Brownian particles in two-dimensions. In addition to interparticle interactions and random bath forces, active Brownian particles are convected along a velocity vector that rotates diffusively with constant amplitude $v_\mathrm{o}$. The specific equations of motion are of the form of Eq.~\ref{Eq:EOM} in an overdamped limit,
\begin{equation}
 \mathbf{\dot{r}}_i =\frac{1}{\gamma} \mathbf{F}_i \left [\mathbf{r}^N \right ] + v_\mathrm{o}\boldsymbol{e}_i +\sqrt{\frac{2 k_\mathrm{B} T}{\gamma} } \boldsymbol{\eta}_i \, ,\quad  {\dot{\theta}}_i = \sqrt{2D_\theta} {\xi}_i
\end{equation}
where $\boldsymbol{e}_i = \{\cos(\theta_i),\sin(\theta_i)\}$ is the unit vector which undergoes diffusive motion due to the random force ${\xi}_i$ and rotational diffusion constant $D_\theta$.  The inter-particle force $\mathbf{F}_i \left [\mathbf{r}^N \right ]$ was derived from the gradient of a WCA potential.\cite{weeks1971role} To characterize diffusive transport, the statistics of single particle displacements was considered, 
\begin{equation}
J_\rho(\tN) = \int_0^{\tN} dt\,  {\dot{r}}_i(t)
\end{equation}
in the long time limit, $\tN \rightarrow \infty$. The distribution of particle displacements was derived by solving the equivalent eigenvalue equation in Eq.~\ref{Eq:Ltilt} for a tagged active Brownian particle. This was possible in an interacting system because Eq.~\ref{Eq:Ltilt} took the form of a Matheiu equation,\cite{Abramowitz1965} with a single unknown $\lambda$-dependent parameter resulting from the closure of a BBGKY-like hierarchy.\cite{Hansen1977} This parameter could be computed numerically from an integral over an empirical pair distribution function. 
 \begin{figure}[t]
\begin{center}
\includegraphics[width=8.5cm]{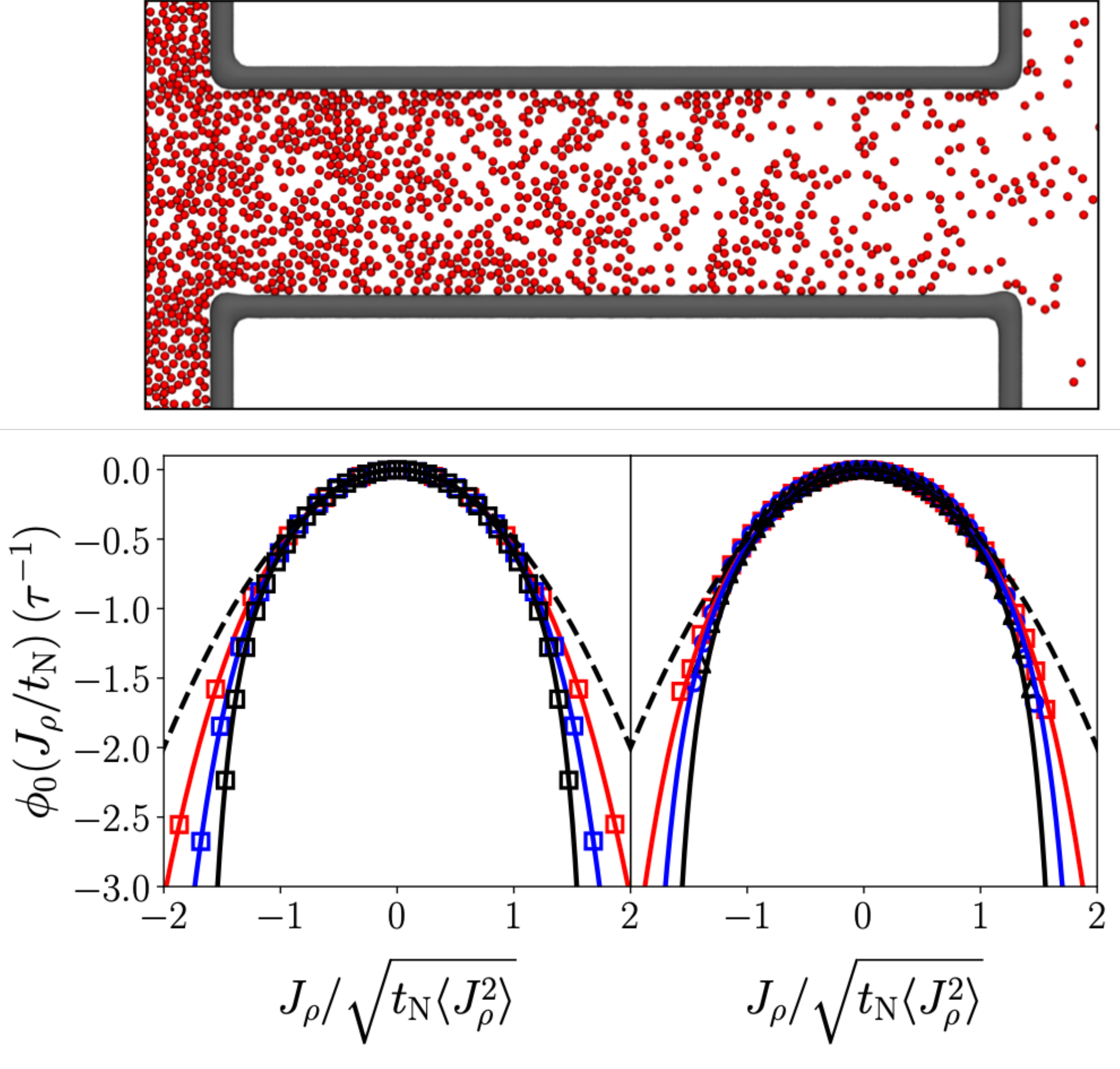}
\caption{The response of a system of active Brownian particles to a density gradient (top) is computable from the rate function of single particle displacement statistics (bottom) for a variety of self-propulsion velocities (left) and densities (right). The symbols denote numerical calculations from the cloning algorithm, solid lines the exact solution of the generalized eigenvalue equation, and the dashed lines a Gaussian rate function with unit variance. Figure adapted from Ref.~\onlinecite{grandpre2018current}.}
\label{Fi:Active}
\end{center} 
\end{figure}

As reported in Ref.~\onlinecite{grandpre2018current}, distribution of single particle displacements was found to be non-Gaussian, except in the limit of passive particles. Example rate functions, $\phi_0(J_\rho/\tN)$, are shown in Fig.~\ref{Fi:Active}, which while parabolic around their means exhibited short tails reflective of suppressed fluctuations around large conditioned displacements. From the variance of the distribution computed at fixed density $\rho$, the self-diffusion coefficient could be determined,
\begin{equation}
\mathcal{D}(\rho) = \lim_{t_\mathrm{N} \rightarrow \infty} \frac{1}{2 \tN} \langle J_\rho^2 \rangle
\end{equation}
which was in excellent agreement with direct estimates from the mean-squared displacement over a range of self-propulsion velocities and densities. Generically, the self-diffusion coefficient decreased modestly with increasing density, and increased significantly with increasing self-propulsion.

The tagged particle current fluctuations encoded by the large deviation rate function provided the response of a hydrodynamic current, $j_\rho$, generated from a slowly varying spatial density, $\rho(r)$. This is analogous to the perspective in Eq.~\ref{Eq:Gaspard}. From the Kramers-Moyal expansion,\cite{Risken1996} $j_\rho$ can be expressed as a gradient expansion
\begin{align}
 j_\rho = -\sum_{n=1}^\infty \frac{(-1)^n}{n ! \tN} \partial_r^{n-1} M^n[\rho(r)] \rho(r)
 \end{align}
 where $M^n[\rho(r)]$ is the local density-dependent $n$th centered moment of the current, $\langle (J_\rho-\langle J_\rho \rangle)^n \rangle$.  To first order at low density, the mass current is linear in the density gradient and is given by Fick's law, $j_\rho \approx -\mathcal{D}(\rho)(\partial \rho/\partial r)$, where $\mathcal{D}(\rho)$ is the proportionality constant relating the current to the gradient. Thus, mass transport is Fickian in that the diffusion constant determines the response of a small density gradient, but nonlinear responses are computable from the density dependence of the current distribution. Direct estimates of gradient diffusivity from the simulation of an initial density gradient were in good agreement with those computed from the rate function. Nonlinear corrections, while relevant for the observed motility induced phase separation in these materials,\cite{cates2015motility} remain to be tested numerically.   

\subsection{Heat transport in low dimensional solids} 
\label{Sec:CNT}
Utilizing our observations that the statistical convergence of linear transport coefficients is accelerated when evaluated from the large deviation function relative to traditional Green-Kubo expressions,\cite{gao2017transport} this approach was applied to study the thermal conduction through low dimensional carbon lattices.\cite{ray2019heat} Heat transport within carbon nanotubes and graphene sheets have received considerable recent attention, due to experimental and simulation reports claiming a violation of Fourier’s law of conduction.\cite{chang2008breakdown,xu2014length,yang2010violation,wang2011non} These reports expand on a large literature of anomalous transport in systems that can be taken macroscopically large in fewer than three-dimensions.\cite{lepri2016heat,cipriani2005anomalous,hurtado2016violation,dhar2008heat}  Experimentally, reports on low dimensional lattices have shown indications of anomalous conductivities, though difficulties extracting definitive values are complicated by boundary effects.  

\begin{figure}[t]
\begin{center}
\includegraphics[width=8.5cm]{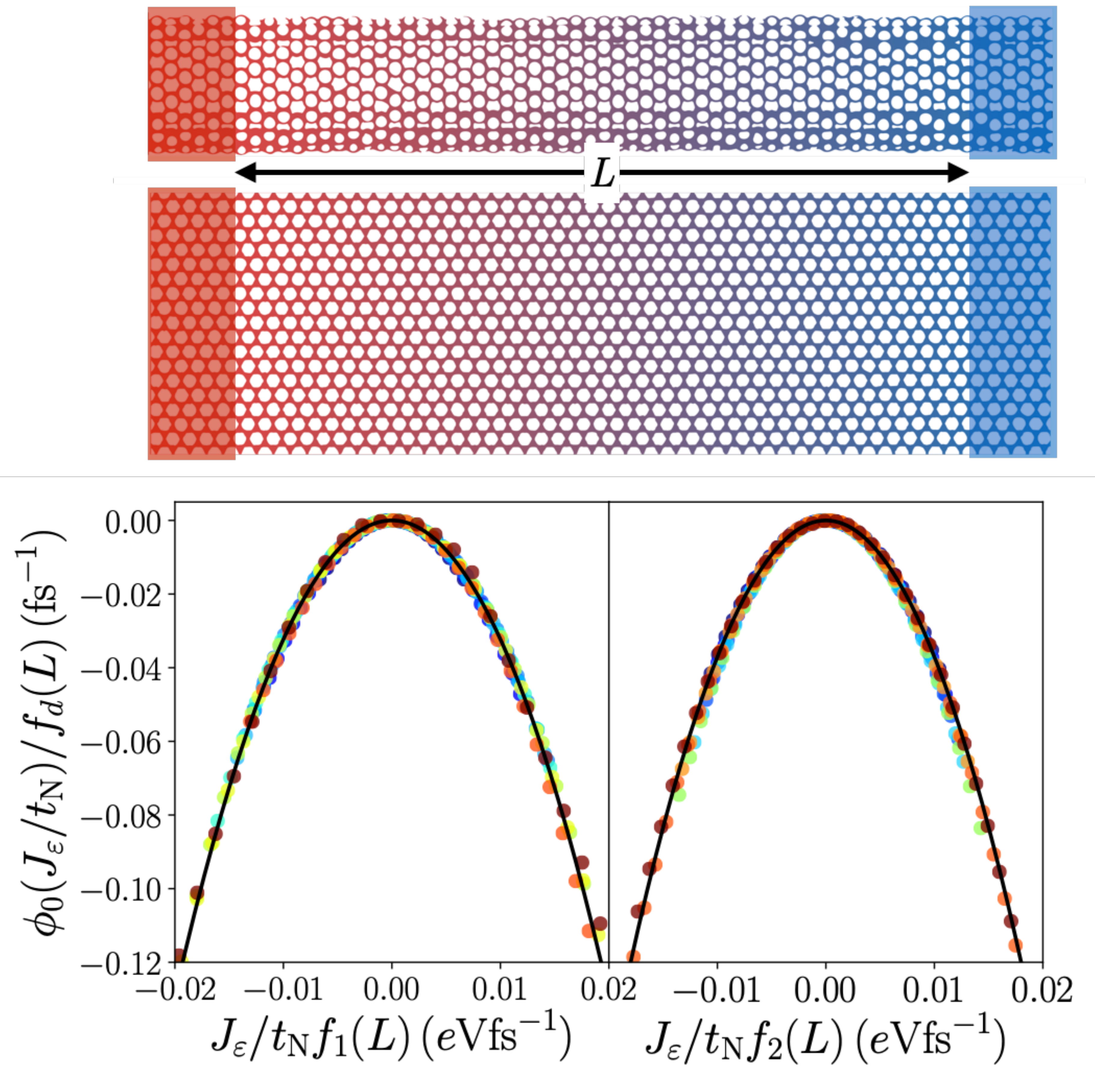}
\caption{Large deviation scaling of the heat current for a nanotube ($d=1$) and graphene sheet ($d=2$). (Top) Illustrates the two geometries in contact with stochastic reservoirs. (Bottom) Rate functions for the heat current for a variety of system sizes, collapsed by scaling functions that asymptotically approach $f_1(L)\sim \sqrt{L/\ell}$ and $f_2(L) \sim L \ln L/\ell$ with characteristic length $\ell$. This figure is adapted from Ref.~\onlinecite{ray2019heat}.}
\label{Fi:4}
\end{center} 
\end{figure}

To understand the mechanism of heat transport in low dimensional carbon lattices, the energy current fluctuations were considered within a nanotube and a graphene sheet. The individual atoms evolved deterministically in the bulk of the material through the solution of Newton's equation of motion, but two stochastic reservoirs at each end imposed a constant temperature through the Langevin equation in Eq.~\ref{Eq:EOM}. The geometry is illustrated in Fig.~\ref{Fi:4}. The atoms interacted through the conservative force described by the gradient of a Tersoff potential parameterized to recover the phonon spectrum of carbon nanostructures.\cite{lindsay2010optimized}

The heat transport was studied by monitoring the energy exchanged with the stochastic reservoirs. Specifically, the energy current through the $k$th reservoir is given by a sum over $N_k$ atoms in that region,
\begin{equation}
\label{Eq:jk}
j_k(t) = \sum_{i \in k}^{N_k} \left [m \dot{\mathbf{v}}_i(t) - \mathbf{F}_i(t)\right ]\cdot \mathbf{v}_i(t)
\end{equation}
and thus the energy exchanged from the $r$th reservoir into the $l$th reservoir over a time $t_\mathrm{N}$ is the integrated current 
\begin{equation}
J_\epsilon =  \int_0^{t_\mathrm{N}} dt\,[ j_l(t)-j_r(t) ]
\end{equation}
in the long time limit, $\tN \rightarrow \infty$. If the system is maintained at thermal equilibrium, with the reservoirs fixed to a common temperature $T$ and separated by a distance $L$, the conductivity is computable from the mean-squared fluctuations of the energy exchanged with the reservoirs,
\begin{eqnarray}
\label{Eq:KapEq}
\kappa_L &=&  \lim_{\Delta T \rightarrow 0} \lim_{t_\mathrm{N} \rightarrow \infty} -\frac{\langle J_\epsilon \rangle_{\Delta T}}{t_\mathrm{N} \Delta T} L\nonumber \\ 
&=& \lim_{t_\mathrm{N} \rightarrow \infty} \frac{\langle  J_\epsilon^2 \rangle_{0}}{2 t_\mathrm{N} \kB T^2}L
\end{eqnarray}
where at long times, for a finite open system, the mean-squared fluctuations are expected to scale linearly with time. This exact expression follows from the definition of $\kappa_L$ as the differential increase of the average heat current with a temperature difference and the stochastic process in Eq.~\ref{Eq:EOM}. This specific form is an example of an Einstein-Helfand moment, equivalent to a Green-Kubo relation.\cite{viscardy2007transport} 

To compute the rate function of heat current fluctuations, the Monte Carlo procedure discussed in Sec.~\ref{Sec:DMC} was employed to importance sample the probability of an integrated current at equilibrium. 
Using a range of $\lambda$’s, a set of $\phi_\lambda(J_\epsilon)$ was related to $\phi_0(J_\epsilon)$ using histogram reweighting,\cite{kumar1992weighted} enabling the construction of $\phi_0(J_\epsilon)$ far into the tails of the distribution. These are shown in Fig.~\ref{Fi:4} for both carbon nanotubes and graphene sheets. Studying a range of system lengths, $\phi_0(J_\epsilon)$ was found to be collapsed with a dimensional-dependent scaling function $f_d(L)$, from which the system-size-dependence of the conductivity was deduced, $\kappa_L\propto f_d(L)/L$. For carbon nanotubes, the thermal conductivity $\kappa_L$ was found to increase as the square root of the length of the nanotube, while for graphene sheets the thermal conductivity was found to increase as the logarithm of the length of the sheet. The particular length dependence and nonlinear temperature profiles place carbon lattices into a universality class with nonlinear lattice models, and suggest that heat transport through carbon nanostructures is better described by a L{\'e}vy walk rather than simple diffusion.\cite{cipriani2005anomalous,delfini2007energy,das2014numerical,liu2014anomalous} However, recent results suggests these anomalous scalings might plateau at even larger lengths than considered in our study.\cite{barbalinardo2021ultrahigh} While this calculation considered linear phenomena, generalizations to heat current rectification have been considered in model systems.\cite{gao2019nonlinear}

\subsection{Ionic conductivity at high fields} 
\label{Sec:NanoF}
The framework presented here allows arbitrary nonlinear transport behavior to be considered on the same footing as traditional linear response. To explore the former, our approach was applied to nonlinear electrokinetic phenomena in ionic solutions. Advances in the fabrication and observation of nanofluidic devices have enabled the study of electrokinetic phenomena on the smallest scales \cite{Faucher_et_al2019,Yang_et_al2018,Radha_et_al2016}. When confined to nanometer dimensions, large thermodynamic gradients can be generated, driving nonlinear responses such as field dependent transport coefficients and nonequilibrium behaviors like current rectification \cite{Siwy_Fulinski2002,Poggioli_et_al2019,Marcotte_et_al2020}. Existing theories for nonlinear conductivities are valid only in the dilute solution regime.\cite{wilson1936theory,demery2016conductivity,donev2019fluctuating} 

\begin{figure}[t]
\begin{center}
\includegraphics[width=8.5cm]{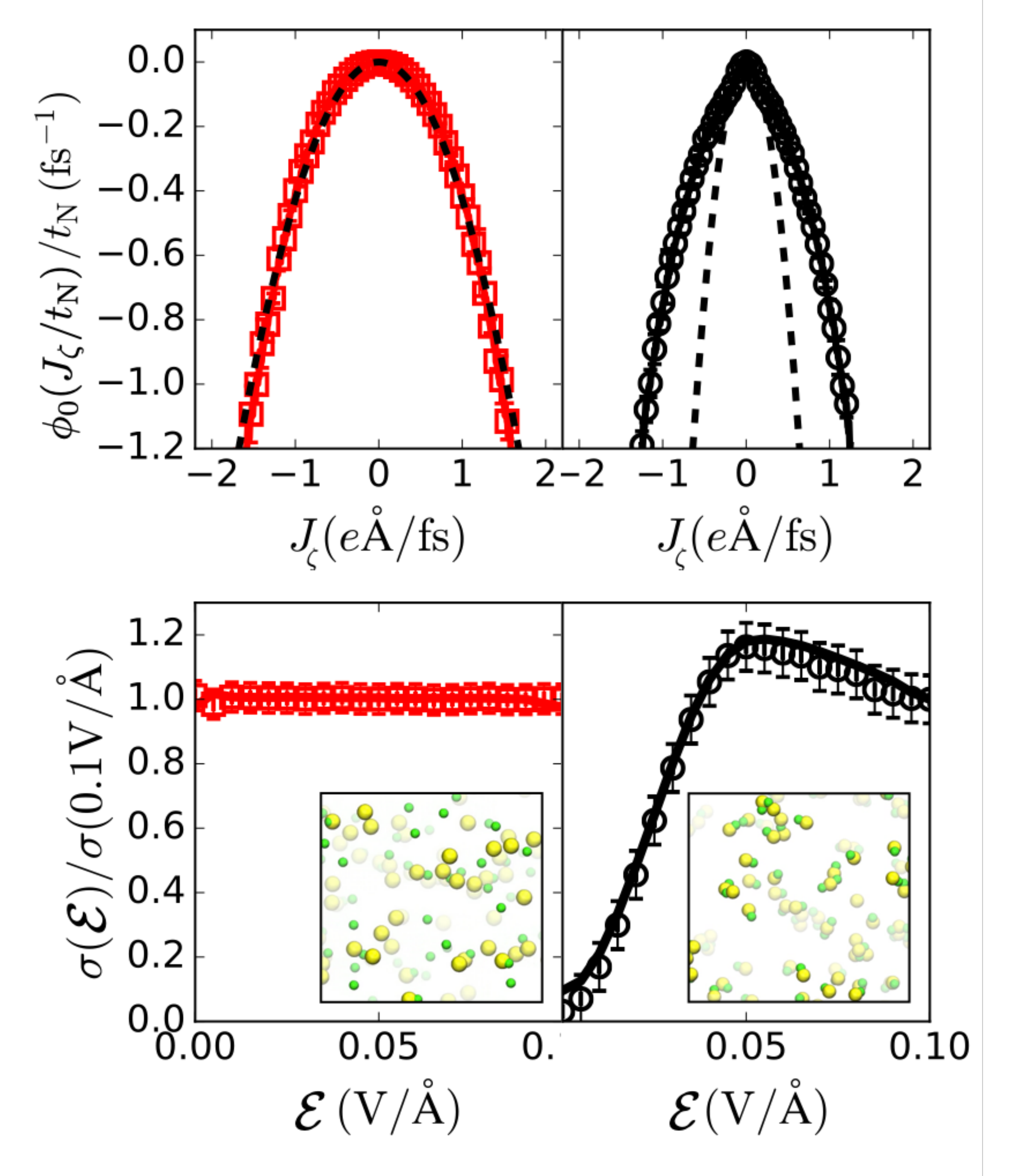}
\caption{Ionic current fluctuations (top) and associated field dependent conductivities (bottom) of a strong (left) and weak (right) electrolyte solution. Insets illustrate characteristic snapshots of the weak and strong electrolyte. Figure adapted from Ref.~\onlinecite{lesnicki2020field}.}
\label{Fi:Ions}
\end{center} 
\end{figure}

The field dependence of the ionic conductivity was studied in strong and weak electrolytes,\cite{lesnicki2020field} developing a contemporary perspective on the so-called Onsager-Wien effect.\cite{onsager1957wien} Initially a monovalent salt was studied in implicit solvents with dielectric constants of 10 and 60 to model a weak and strong electrolyte, respectively.  The ions evolved with Eq.~\ref{Eq:EOM}, with frictions chosen to recover the self diffusion coefficients in the dilute limit. In order to predict the field dependent ionic conductivity from equilibrium fluctuations, knowledge of both the ionic current
\begin{equation}
J_\zeta  = \sum_i  \int_0^{\tN} dt \, z_i {v}_i
\end{equation}
where $z_i$ is the charge on ion $i$, and its time reversal symmetric counter part, and the dynamical activity,
\begin{equation}
Q_\zeta  = \sum_i  \int_0^{\tN} dt \, \frac{z_i}{\gamma_i} \left (m_i {\dot{v}}_i - F_i \left [\mathbf{r}^N \right ]\right)
\end{equation}
which is a difference between momentum flux and intermolecular force weighted by the charge and friction is needed. Using nonequilibrium ensemble reweighting, the joint rate function $\phi_0(J_\zeta,Q_\zeta)$ was computed far into its tails. This reweighting procedure is made possible by the relationship given in Eq.~\ref{Eq:Gao}, where typical fluctuations of $J_\zeta$ and $Q_\zeta$ for simulations under finite applied fields can be used to reconstruct rare fluctuations in the absence of a field. The marginalization of the joint distribution constructed from a series of nonequilibrium simulations onto the current is shown in Fig.~\ref{Fi:Ions}. For the strong electrolyte, the current fluctuations were found to be incredibly Gaussian. For the weak electrolyte, locally Gaussian fluctuations around its mean broaden significantly into fat tails. The tails were well described by a second Gaussian with larger variance, manifesting the suppression of current fluctuations when ions are paired in the weak electrolyte and its enhancement when they dissociate upon conditioning on a large current. 

With the joint rate function, the differential conductivity, $\sigma(\mathcal{E})$, was computed as a continuous function of an applied field $\mathcal{E}$. This follows from the definition of $\sigma(\mathcal{E})$, as the differential change in the the current density with applied field
\begin{eqnarray}
\sigma(\mathcal{E}) &=& \frac{1}{\tN V} \frac{d\langle J_\zeta \rangle_\mathcal{E} }{d \mathcal{E}}  \\
&=& \frac{1}{2 \tN \kB T V} \langle (\delta J^2_\zeta+\delta J_\zeta \delta Q_\zeta) e^{\beta (J_\zeta+Q_\zeta)\mathcal{E}/2}  \rangle_0 \nonumber
\end{eqnarray}
where $V$ is the system volume and the long time limit, $\tN \rightarrow \infty$ is taken to evolve a nonequilibrium steady-state. While the first line is a definition, the second line employs the nonequilibrium ensemble reweighting relation. 

Figure ~\ref{Fi:Ions} illustrates the differential conductivity computed in this way, which is in good agreement with that evaluated  from a numerical derivative of the current-field relationship; however, the latter is statistically much more difficult to converge. As anticipated from the Gaussian current statistics, the strong electrolyte exhibits a field-independent conductivity equal to its Nernst-Einstein value. By contrast, the weak electrolyte, with its marked non-Gaussian current statistics, exhibits a strongly field dependent conductivity. For the weak electrolyte, an initially low value of $\sigma(\mathcal{E})$ increases, exhibits a small maximum, before plateauing to its Nernst-Einstein limit. The maximum reflects the increased fluctuations at fields just strong enough to dissociate ion pairs. Opposing these enhanced current fluctuations are negative correlations between $j$ and $q$, which reflect ionic relaxation dynamics whereby electric fields generated by distortions of the ionic cloud around an ion are anti-correlated with displacements of the ion in the direction of the external field.  Extensions of this analysis to explicit solvent models has been recently undertaken,\cite{lesnicki2021molecular} and their implications for simulations with explicit nanoconfinement remains to be studied.\cite{pean2015confinement,simonnin2018mineral,palmer2020correlation}

\section{beyond}
The perspective articulated here illustrates how to leverage formal advances in nonequilibrium statistical mechanics and novel numerical techniques to address contemporary problems in nanoscale transport. While significant strides have been made recently in applying large deviation theory to molecular systems driven far from equilibrium, there are clearly outstanding questions. 

First, there are technical issues associated with the appropriate equation of motion to describe molecular systems away from equilibrium. Near equilibrium, ensemble equivalence\cite{lebowitz1967ensemble} requires linear response functions to be equal whether they are propagated under Newtonian or stochastic equations of motion, provided the characteristic timescale of the bath is large. Away from equilibrium, however, nonlinear response functions depend on the details of the equation of motion. The constant supply of energy through an applied field requires a means to dissipate that energy in order to evolve a nonequilibrium steady state, so absent explicit boundaries a thermostat of some sort must be used. While in some cases the details of the thermostat can be motivated physically, extending the nonlinear response formalism and sampling algorithms discussed here to non-Markovian and deterministic equations of motion would provide for alternative modeling choices and further generality. Typically physically derived non-Markovian equations can be embedded to yield Markovian models in larger phase spaces.\cite{bao2005non,baczewski2013numerical} Exploring connections to other response formalisms employed with deterministic thermostats would undoubtedly be fruitful.\cite{morriss1985isothermal,morriss1987application,van2004extended} Similarly, we have focused entirely on classical systems, but extending this perspective to quantum mechanical transport problems would undoubtedly yield novel insights.  In weak coupling regimes this is likely possible; however, away from these regimes it is uncertain.\cite{levy2020response,konopik2019quantum} 

Second, there are issues of how to translate connections between underlying microscopic dynamics and mesoscopic behavior into novel design principles. The response theories relate particular dynamical correlations to emergent transport phenomena, and have been successfully used to explain experimental observations. However, inverting that relationship, and rationally designing a molecular system with a target emergent response is  difficult. A potential route to this inverse design is to view Eq.~\ref{Eq:Var} as a cost function to be optimized. Its interpretation is clear, with the added drift being the smallest change to an equilibrium system to make a specific current response typical within its steady state.  Indeed, this insight has already been used in the context of nonequilibrium self-assembly,\cite{das2021variational} and in the design of flow fields for tracer particles.\cite{pineros2021inverse} A number of outstanding challenges in renewable energy, separations, and computation could be solved provided a nonequilibrium inverse design principle.\cite{fleming2008grand} Within the perspective here, this seems possible. 

\section{ACKNOWLEDGMENTS} 
The authors would like to thank 
Garnet Chan, Avishek Das, Juan Garrahan, Phillip Geissler, Trevor GrandPre, Robert Jack, Kranthi Mandadapu, Benjamin Rotenberg and Hugo Touchette for useful discussions. This material is based upon work supported by NSF Grant CHE-1954580.
%\pagebreak

\section{Author Contribution Statement}
D. T. L, C. Y. G and A. R. P wrote the manuscript, designed and performed the research. 

\section{References}
\bibliography{ref_arp}

\end{document}